\documentclass{article}

\usepackage{arxiv}

\usepackage[utf8]{inputenc} 
\usepackage[T1]{fontenc}    
\usepackage{hyperref}       
\usepackage{url}            
\usepackage{booktabs}       
\usepackage{amsfonts}       
\usepackage{nicefrac}       
\usepackage{microtype}      
\usepackage{lipsum}

\usepackage{stackengine,graphicx, caption}
\usepackage{multicol}

\newenvironment{Figure}
  {\par\medskip\noindent\minipage{\linewidth}}
  {\endminipage\par\medskip}
 
\title{MitoDet: Simple and robust mitosis detection}

\author{
  Jakob Dexl, 
  Michaela Benz, Volker Bruns, Petr Kuritcyn, Thomas Wittenberg\\
  Fraunhofer Institute for Integrated Circuits IIS\\
  Friedrich-Alexander-Universität Erlangen-Nürnberg (FAU)\\
  \texttt{jakob.dexl@fau.de} \\
}

\begin{document}
\maketitle

\begin{multicols}{2}[]
\begin{abstract}
Mitotic figure detection is a challenging task in digital pathology that has a direct impact on therapeutic decisions. While automated methods often achieve acceptable results under laboratory conditions, they frequently fail in the clinical deployment phase. This problem can be mainly attributed to a phenomenon called domain shift. An important source of a domain shift is introduced by different microscopes and their camera systems, which noticeably change the colour representation of digitized images. In this method description, we present our submitted algorithm for the Mitosis Domain Generalization Challenge \cite{aubreville_mitosis_2021}, which employs a RetinaNet \cite{lin_focal_2017} trained with strong data augmentation and achieves an F1 score of 0.7138 on the preliminary test set.
\end{abstract}
\keywords{Mitosis detection \and Domain generalization \and Digital pathology}
\section{Methods}
Motivated by recent data-centric approaches we use a RetinaNet \cite{lin_focal_2017} trained with strong data augmentation to enforce prediction consistency.

\subsection{Dataset}
We use the publicly available Mitosis Domain Generalization Challenge (MIDOG) dataset \cite{aubreville_mitosis_2021}. The data consists of 200 Whole Slide Images (WSIs) from hematoxylin and eosin (HE) stained breast cancer cases. Furthermore, the dataset can be divided into subsets of 50 images, which were acquired and digitized with four different scanners (Aperio ScanScope CS2, Hamamatsu S360, Hamamatsu XR NanoZoomer 2.0, Leica GT450). For three scanners annotations for mitotic figures and hard negatives (imposters) are provided. The disclosed preliminary and final test sets contain samples of two known scanners and two unknown ones. 

\subsection{Model}
Our object detection algorithm consists of a RetinaNet \cite{lin_focal_2017} with an EfficientNet B0 \cite{tan_efficientnet_2019} backbone. The backbone is initialized with state of the art ImageNet weights, which were trained using RandAugment \cite{cubuk_randaugment_2020} and Noisy Student \cite{xie_self-training_2020}. 
We did not change the feature pyramid and used all five pyramid levels. 
The network's heads consist of four layers with a channel size of 128. Anchor ratios are set to one while the differential evolution search algorithm introduced by \cite{zlocha_improving_2019} is employed to determine three anchor scales (0.781, 1.435, 1.578). 

\subsection{Domain generalization through augmentation}
Our main method to approach domain generalization is data augmentation. Data-driven approaches such as RandAugment \cite{cubuk_randaugment_2020} have been proven to increase model robustness and have been used in recent state of the art models. 
\begin{Figure}
 \centering
 \includegraphics[width=\textwidth, trim=0pt 75pt 0pt 75pt,clip]{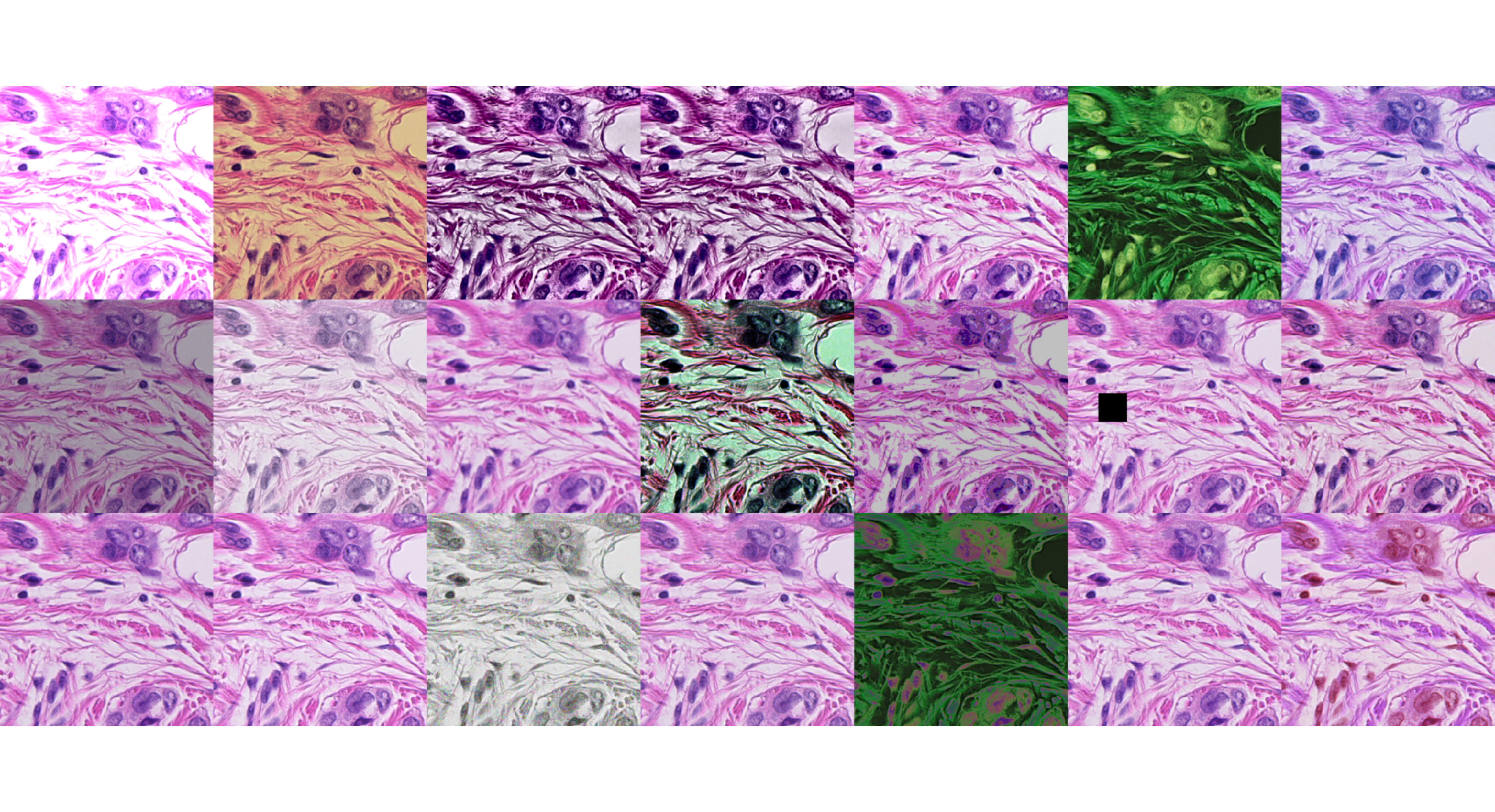}
 \captionof{figure}{Used augmentations with different strengths.}
\end{Figure}
Inspired by Trivial Augment \cite{muller_trivialaugment_2021} a very simple random augmentation strategy is used, where a single augmentation is applied to each image. The augmentations are drawn uniformly from a set of color, noise and special transformations while the augmentation strength is random to some defined degree. The pool of augmentations consists of color jitter, HE \cite{tellez_whole-slide_2018}, fancy PCA, hue, saturation, equalize, random contrast, auto-contrast, contrast limited adaptive histogram equalization (CLAHE), solarize, solarize-add, sharpness, Gaussian blur, posterize, cutout, ISO noise, JPEG compression artefacts, pixel-wise channel shuffle and Gaussian noise. In addition, every image is randomly flipped and RGB channels are randomly shuffled. 

\subsection{Training and evaluation}
For experimentation, we divide the dataset into five folds with three training, one validation and one test split for each scanner (test splits are added to the train set for submissions). During the training phase, we uniformly sample the images of the train set and randomly select a mitotic figure or an imposter annotation. A patch with a size of 448 pixels is randomly cropped around the selected annotation similar to \cite{marzahl_deep_2020}. The RetinaNet is trained for 100 pseudo epochs with a batch size of 16 using the super-convergence scheme \cite{smith_super-convergence_2019}. Adam optimizer with a maximum learning rate of 1e-4 is used. The best models are selected based on the lowest validation loss. After the training phase, we combine the training and validation set and optimize the model's confidence threshold with respect to the best F1 score. During inference, incoming WSIs are tiled into overlapping patches of 448 pixels. All models are trained and tested using an Nvidia GeForce RTX 3060 with 12GB GPU RAM. 

\section{Results}
For the final submission, we only use labelled data to train a single RetinaNet with the proposed data augmentation strategy. This method achieves an F1 score of 0.7138 on the preliminary test set of the MIDOG challenge.

\section{Discussion}
Overall, we are able to generalize better across multiple scanner domains with strong data augmentation.
The magnitude at which such simple transformations improve generalization at no cost of inference speed is higher than expected. Even models trained with only one scanner reach similar results on our test split, showing only a small performance drop. 
In the following, we will lay out unsuccessful attempts to improve the quality further.
One major issue was the model selection based on the validation loss. The models were not capable of overfitting the data, assumingly due to the sampling and the strong data augmentation, models ended up in an equilibrium mode where performance improvements were wiggling between the different scanners back and forth. Because of that, the representation shift metric proposed by Stacke et al. \cite{stacke_measuring_2021} was tested. It was applied to the three convolutional layers, which flow into the feature pyramid, but was found to not help the model selection process. 
Another strategy was a dual-stage attempt with a verification net proposed by Li et al. \cite{li_deepmitosis_2018}. The network was trained on the predicted patches of the first stage using the same augmentation and in addition a Gradient Reversal Layer \cite{ganin_domain-adversarial_2016} to remove even the last bits of scanner dependent information. Unfortunately, this resulted in a performance drop of 12.1\% on the preliminary dataset. 
Finally, the choice of using an EfficientNet originated from the attempt to incorporate the unlabeled data using a self-supervised Student-Teacher learning procedure based on the STAC framework \cite{sohn_simple_2020}. While increasing the performance on our test split, this resulted in a small performance drop of 1\% on the preliminary dataset. One problem was that producing pseudo labels with a high confidence threshold resulted in very few labelled samples while self-training reportedly needs a huge amount of pseudo labelled data to make use of it. A second problem arises with false positive pseudo labels. We used a labelled scanner to check the number of wrong labels incorporated in the pseudo labels and found that for mitotic figures pseudo labels were mainly correct while hard negatives actually included a lot of mitotic figures. This probably led to more confusion than having a positive effect.

\section*{Acknowledgments}
This work was supported by the Bavarian Ministry of Economic Affairs, Regional Develop-
ment and Energy through the Center for Analytics – Data – Applications (ADA-Center)
within ”BAYERN DIGITAL II” and by the BMBF (16FMD01K, 16FMD02 and 16FMD03).

\end{multicols}
\end{document}